\documentclass[amssymb,twocolumn,aps,pra,superscriptaddress,10pt]{revtex4}
\usepackage{centernot}
\usepackage{algorithm}
\usepackage{algpseudocode}

\usepackage{amsmath,amssymb,bm}
\usepackage{graphics}
\usepackage{epstopdf}
\usepackage{epsfig}
\usepackage{graphicx}
\usepackage{eqnarray,amsmath}
\usepackage{color}
\usepackage{comment}
\usepackage{xcolor}

\newcommand{\mean}[1]{\left\langle#1\right\rangle}
\newcommand{\er}{Erd\H{o}s-R\'enyi }
\newcommand{\ba}{Barab{\'a}si-Albert }

\begin{document}

\linespread{1}

\title{Robustness of Noisy Quantum Networks}

\author{Bruno~C. Coutinho}
\affiliation{Instituto de Telecomunica\c{c}\~{o}es, Physics of Information and Quantum Technologies Group, Portugal}

\author{William J. Munro}\affiliation{NTT Basic Research Laboratories \& NTT Research Center for Theoretical Quantum Physics, NTT Corporation, 3-1 Morinosato-Wakamiya, Atsugi-shi, Kanagawa 243-0198, Japan}
\affiliation{National Institute of Informatics, 2-1-2 Hitotsubashi, Chiyoda-ku, Tokyo 101-8430, Japan}

\author{Kae Nemoto}
\affiliation{National Institute of Informatics, 2-1-2 Hitotsubashi, Chiyoda-ku, Tokyo 101-8430, Japan}

\author{Yasser Omar}
\affiliation{Instituto de Telecomunica\c{c}\~{o}es, Physics of Information and Quantum Technologies Group, Portugal}
\affiliation{Instituto Superior T\'ecnico, Universidade de Lisboa, Portugal}

\date{\today}

\begin{abstract}{Quantum networks are a new paradigm of complex networks, allowing us to harness networked quantum technologies and to develop a quantum internet. But how robust is a quantum network when its links and nodes start failing? We show that quantum networks based on typical noisy quantum-repeater nodes are prone to discontinuous phase transitions with respect to the random loss of operating links and nodes, abruptly compromising the connectivity of the network, and thus significantly limiting the reach of its operation. Furthermore, we determine the critical quantum-repeater efficiency necessary to avoid this catastrophic loss of connectivity as a function of the network topology, the network size, and the distribution of entanglement in the network. In particular, our results indicate that a scale-free topology is a crucial design principle to establish a robust large-scale quantum internet.}
\end{abstract}
\maketitle

\clearpage


\section{Introduction} 

 Quantum networks are a new paradigm of networks where the links and/or the nodes obey the laws of quantum physics \cite{Kimble2008,Caleffi2018,vanmeter2014}. Namely, the quantum links can be quantum correlations \cite{citeulike:7231538}, quantum couplings or dynamics \cite{inlek2017multispecies,PhysRevLett.81.5932}, or quantum causal relations \cite{brukner2014quantum}. And quantum nodes can be any system with quantum degrees of freedom. The nascent field of complex quantum networks~\cite{PhysRevLett.119.220503,citeulike:7231538,PhysRevLett.124.210501,Pirandola2019_1,Pirandola2019_2,PhysRevA.101.052315,PhysRevLett.116.100501,PhysRevX.4.041012,Biamonte2019,PhysRevA.77.022308,PhysRevLett.103.240503,PhysRevA.83.032319,PRXQuantum.2.010304} is motivated both by the fundamental interest of understanding the nature and the properties of this new object, as well as by the applied perspective of developing networked quantum technologies to fully harness their potential and their reach, namely for quantum-secure communications \cite{gisin2007quantum,Gisin2002}, for quantum-accelerated computation \cite{bennett2000quantum,arute2019quantum,Zhong1460}, for quantum-enhanced sensing and metrology \cite{RevModPhys.89.035002,caves1982,Giovannetti1330}, and, overall, for the development of a future quantum Internet \cite{Kimble2008}. However, quantum states and quantum systems are notably vulnerable to noise in general. But how does this translate to the network realm, i.e.\ how robust are noisy quantum networks, and how is that robustness affected by the underlying graph? And how does it compare to the robustness of classical networks, which typically evolve, to non-trivial network topologies~\cite{dorogovtsev2013evolution,barabasi2016network}, such as scale-free properties, topologies that are known to maintain their functionality against random failures \cite{barabasi2016network,PhysRevLett.85.4626,chen2017robustness}?


\section{Quantum Networks} 

\begin{figure*}
\centering
\includegraphics[width=1\textwidth]{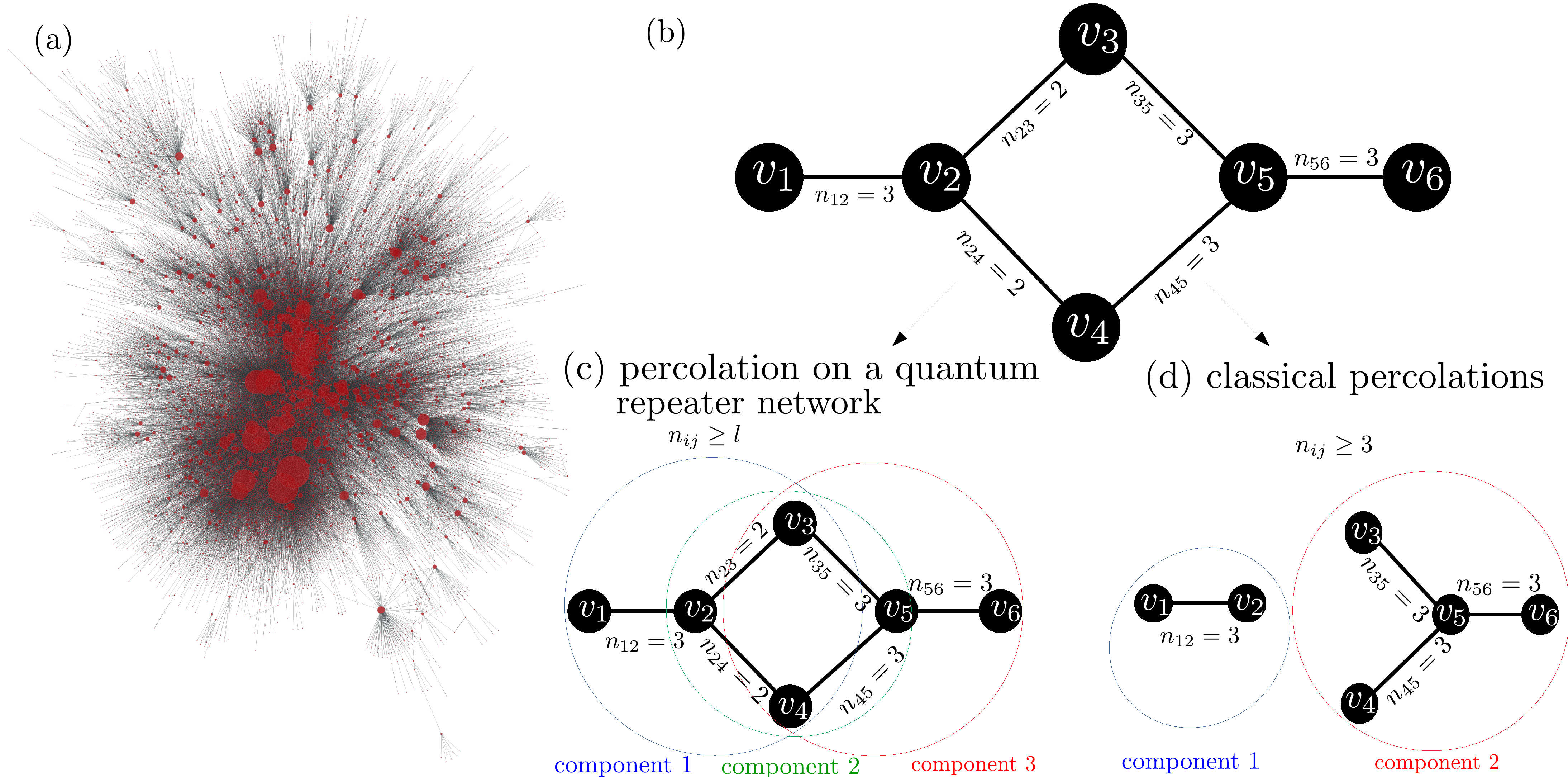}
\caption{{{\bf Complex quantum networks} --- Depicted in (a) is a snapshot of the structure of  Internet (at the level of autonomous systems ~\cite{internetdata}) clearly showing the scale free properties of this complex network. This snapshot could in principle belong to a future quantum internet \cite{Kimble2008,wehner2018quantum} which will however operate on different network principles. These differences can be seen even at the small scale. In (b)  a small scale quantum repeater network is shown, indicating how the connected components can intersect each other, in stark contrast to what is  observed in a classical network. Here each node is represented by a black dot and  the links by black lines. $n_{ij}$ represents the number of entangled pairs associated with each link $e_{ij}$, which is chosen for this illustration to scale as $r(l)=l$. Two nodes $v_i$ and $v_j$ are connected at a distance $l$ if there is a path between them such that for all links in that path satisfy the condition $n_{ij}\geq l$. As an example (c) illustrates a quantum network where one can only connect two nodes if $n_{ij}\geq l$. The connected components clearly intersect each other. In contrast (d) illustrates a classical network where links can only be used to connect two nodes if $n_{ij}\geq 3$. In this case the connected components do not intersect each other. 
\label{fig1}}}
\end{figure*}

Networks are a set of nodes and links, where each link is a pair of nodes. This naturally includes complex networks  \cite{barabasi2016network,newman2018networks,mezard2009information} such as the current classical Internet~\cite{internetdata}, a snapshot of which is presented in Fig.~\ref{fig1}(a). With the goal of investigating a quantum Internet, we consider quantum networks where the links correspond to entangled pairs of qubits, each lying in a different node, and a node contains one qubit for each of its neighbors. Now, imagine we want to realise a quantum operation, e.g. computing, communication, or metrology, between two distant nodes of a quantum network: how can we connect them, i.e.\ how can we establish entanglement between them, with a certain target fidelity $F_{\rm target}$, given the existing quantum correlations in the quantum network?

 Let us consider the general scenario where there are $N_{ij}$ noisy Bell pairs with  fidelity $F_{\rm initial}$ connecting nodes  $v_i$ and $v_j$. If necessary, these noisy Bell pairs can be purified to yield $n_{ij} =  N_{ij}/N_{\rm ft}$ pairs exceeding a given targe fidelity $F_{\rm target}$ (where  $N_{\rm ft}$ is the number of initial pairs necessary to generate one $F_{\rm target}$ pair) \cite{munro2015inside,Bennett1996}. Next, entanglement swapping between link $v_i$ \& $v_j$ and link $v_j$ \& $v_k$ consumes those Bell pairs to create a longer-range entangled pair between nodes $v_i$ \& $v_k$ with fidelity $\sim F^2_{\rm target}$ \cite{Bennett1993,PhysRevLett.71.4287}. That drop in fidelity  means multiple pairs need to be available for entanglement purification to return the target fidelity $F_{\rm target}$  (again consuming more pairs). These entanglement swapping and purification operations continue until we have connected  the nodes/users who want to communicate in the network \cite{munro2015inside,Muralidharan2016}. A critical question that arises is the resource consumption in such an approach. Fortunately, it is well known that resources required for the first-generation quantum-repeater network scale polynomially with the number of links $l$ needed to connect the source node Alice and  Bob.  To the leading order in this polynomial, we can define \cite{Muralidharan2016},
\begin{eqnarray}
&R(l)=l^{\alpha+1}=r(l)l
\end{eqnarray}
as being the number of entangled qubit pairs in the entire chain necessary to create the connected entangled qubit pair with the desired fidelity $F_{\rm target}$. Further $r(l)=l^{\alpha}$ is the number of entangled qubit pairs per link necessary to create that connected entangled qubit pair. Here $\alpha$  represents the efficiency of the protocol which depends heavily on the experimental apparatus used for the repeater scheme and the noise present in it. As an example, the Werner state and rank 2 Bell state have $\alpha \sim 2, 1$ respectively for a target fidelity of $0.99$ using the ``purification and swap" repeater architectures with perfect local gates.  This compares with  $\alpha=0$ for noiseless Bell states.

 The study of the connectivity of a quantum repeater network requires the introduction of two types of connection between nodes: that we are going to call functional and structural connectivity. 
 Functional connectivity in the quantum regime is the situation where a connection between the two nodes can be established with fidelity $F_{\rm target}$. Structural connectivity on the other-hand refers to the situation where a connection, since there a path connecting the two nodes, but not necessary with fidelity $F_{\rm target}$. We will illustrate these two concepts in Fig.~\ref{fig1} where the nodes $v_1$ \& $v_3$ (and $v_3$ \& $v_5$) can individually establish connections with sufficient fidelity $F_{\rm target}$ (functionally connected), but $v_1$ \& $v_5$ while connected can not (structurally connected). This means $v_1$ does not belong to the same ``functionally" connected component as $v_5$ making it impossible to establish a connection between them with the required fidelity. 


{Standard Bernoulli percolation \cite{Sahini1984,dorogovtsev2008critical,barabasi2016network,newman2018networks}, a widely used technique to explore to the robustness of classical networks\cite{Sahini1984,dorogovtsev2008critical,barabasi2016network}, cannot be used in these quantum scenarios due to the quality of service $F_{\rm target}$ requirement. Although in Bernoulli percolation theory, finding the largest connected component of a network is a computationally easy problem to solve~\cite{newman2018networks}, the largest functional component of any network is an NP-hard problem (see Appendix.\ref{appendix1} for details)}. Our model is more manageable if one considers the case where all links perform the same number of rounds of purification subject to the constraint that $n_{ij}\geq n_{\rm fixed}$ (with $n_{\rm fixed}$ being the number of entangled pairs necessary to successfully perform the purification). Interestingly $n_{\rm fixed}$ can be associated with a distance $l_{\rm fixed}=n_{\rm fixed}^{1/\alpha}$, meaning this link can be functionally used in paths of length $l_{\rm fixed}$ or less.

\begin{figure*}
\centering
\includegraphics[width=0.8\textwidth]{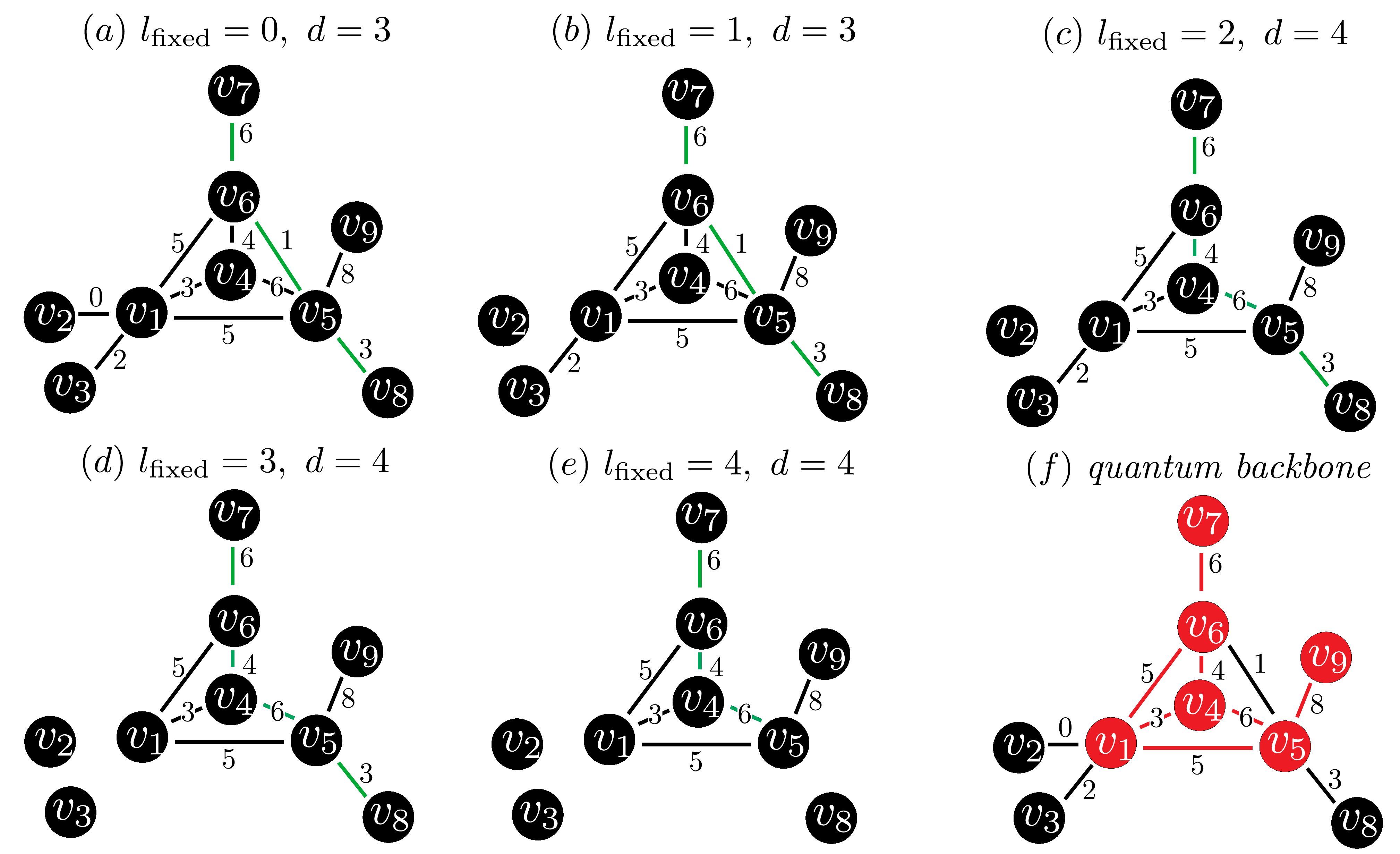}
\caption{ {\bf Quantum network backbone} --- Illustration of how the  {\it quantum backbone} can be computed where the numbers written in each link represents the number of entangled pairs contained in it. We begin in (a) with a $d=3$ network and $l_{\rm fixed}=0$ meaning none of the links in the network need to be removed. As $l_{\rm fixed}$ is increased from 1 (b) to 2 (c) we see that we have removed links with $n_{ij}< 2$. Further in (c) we notice that  the distance of the network increases from $d=3$ to $d=4$ due to the $v_5$ - $v_6$ link being removed in (b) due to lack of resources. Then in (d-e) we continue to remove various links until we reach $n_{ij}<l_{\rm fixed}^{\alpha}$. We are now left in (e) with the largest connected network component, termed our {\it{quantum backbone}}. Finally in (f) we superimpose this  {\it{quantum backbone}} (shown in red) onto the entire network. The size of the backbone (number of nodes in the largest connected component of the backbone) is the number of red nodes nodes in $(f)$, namely $6$. \label{fig2}}
\end{figure*}

Our concept of quantum functional connection naturally suggests that we should choose $n_{\rm fixed}$ as large as possible as it enables nodes on paths of length $l \leq l_{\rm fixed}$ to establish Bell pairs with our required fidelity $F_{\rm target}$. However increasing $l \leq l_{\rm fixed}$ reduces the  probability of a given link have the required number of entangled pairs. Thus there is an important trade off to consider. If the number of pairs distributed between nodes can be expressed as a function $g(n)$, then the probability that a link has at least $n_{\rm fixed}$ pairs is simply
\begin{equation}
p_{\rm fixed}(l_{\rm fixed})=\int_{n_{\rm fixed}}^{+\infty} g(n) dn
\end{equation}
which indicates that for $n_{\rm fixed}$ larger than a certain value, most links are removed from the network and there is no giant component in the network.  Instead one needs to find the largest value of $n_{\rm fixed}$, such that $l_{\rm fixed}\leq d$ with $d$ being the diameter of the network~\cite{barabasi2016network,newman2018networks} (the largest distance between any two nodes in the connected component). In that case, the two nodes (associated with the  connected component) are able to establish a functional connection (see Fig.~\ref{fig2}). This set of nodes is termed the {\it{backbone}} and serves two purposes: first it can be used as a measure of the connectivity of the network, and second for large networks they will behave similarly to classical communication networks. Routing developed for classical networks~\cite{dijkstra1959note,kurosecomputer,TanenbaumWetherall11} should be sufficient (although not necessarily optimal) to find a good path to connect our two nodes with fidelity $F_{\rm target}$~. One should vary  $l_{\rm fixed}$ for any change in the network (like the removal of nodes and links) in order to maximize the size of the {\it backbone}. This however is not trivial to achieve and so it is useful to define two useful quantities:
\begin{itemize}
\item First is $L(x)$ which is the distance $l_{\rm fixed}$ written as a function of $x\equiv-\ln(p_{\rm fixed})$ incorporating the form and efficiency of the quantum repeater protocol used (including resources used and their distribution), 
\item Second $D(x+y)$, Here  the diameter of the largest connected component of the network parameterized by $p=e^{-(x+y)}$ which is the probability that we both have sufficient pairs to create our link (given by $p_x=e^{-x}$) and that their have not been any random failures given by ($p_y=e^{-y}$). Next the diameter and size of the largest connected component $D(x+y)$  can be evaluated using the normal percolation tools \cite{newman2018networks}. 
Now one wants to establish the repeater network with just enough resources to reach the diameter of the largest connected network component ($l_{\rm fixed}=d$ ideally). In that a case we want to find values of $x=x_0$ such that
\end{itemize}
\begin{equation}
L(x_0)=\text{D}(x_0+y),
\label{eqdiamter_3}
\end{equation}
At this $x_0$ point we have the minimal number of resources required to reach the diameter of the networks largest connected component. After we find $x_0$ the size of the {\it backbone}, can be easily computed as the number of nodes in largest connected component of the network composed of only links with sufficient pairs to create our link and that their have not been any random failures. In the example of Fig.~\ref{fig2}, since there are no random failures the size of the {\it backbone} is 6. Adding random failures just means that we are starting with a network where some of the links have been already removed.  Our approach is slightly simplistic in that we have only considered links (loss of nodes can also be incorporated).
We are now at the stage where we can explore actual networks. 

\section{Quantum \er  Networks} 

\begin{figure*}
\centering
\includegraphics[width=0.8\textwidth]{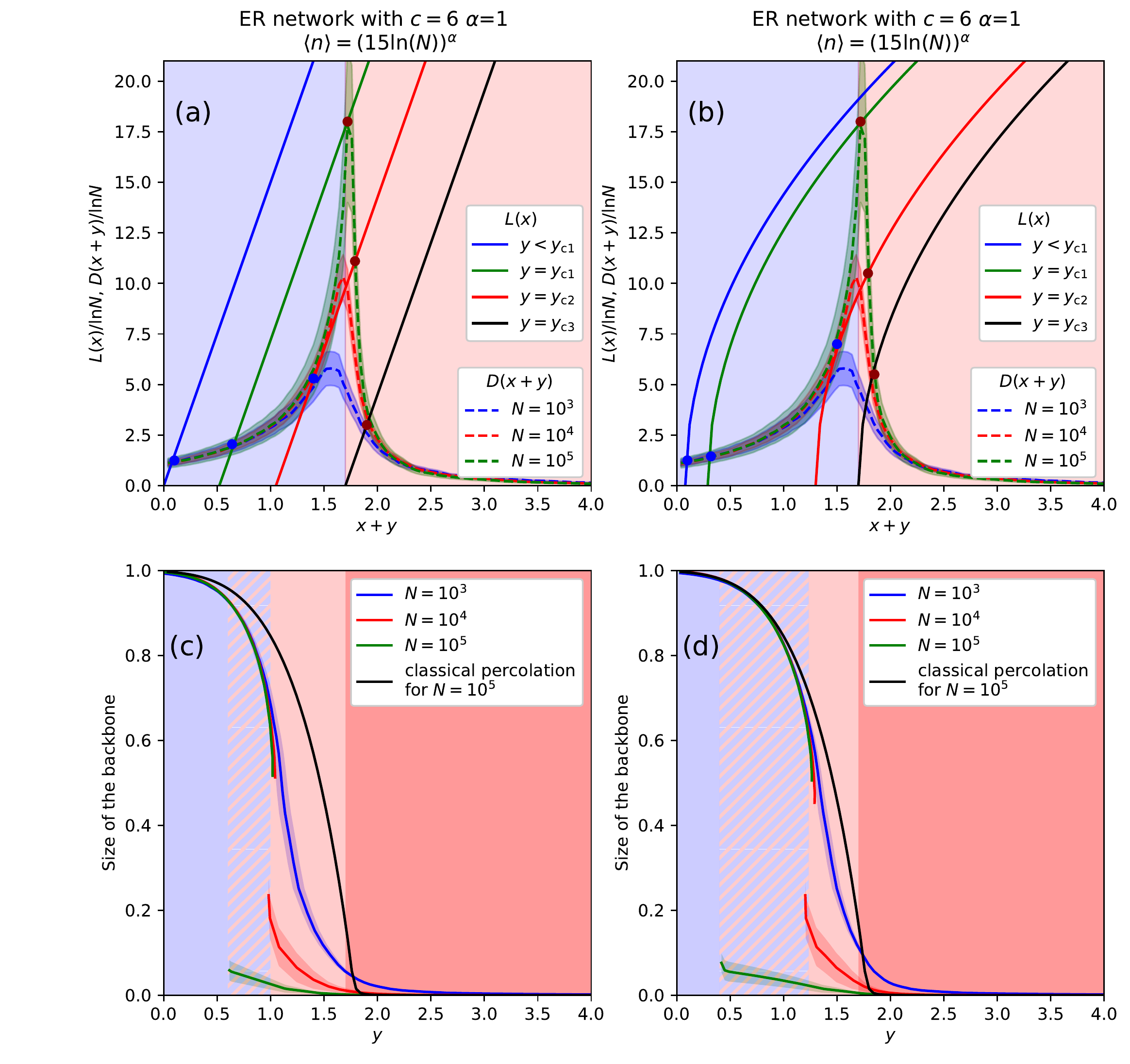}
\caption{{{\bf Robustness of a quantum Erd\H{o}s-R\'enyi   network} --- Exploration of  bond percolation on a quantum \er   network  with average degree  $c=6$  where the number of entangled pairs in each link follows an exponential distribution with mean number $\langle n \rangle$. We plot $L(x)$ and $D(x+y)$ versus $x+y$ for $\langle n \rangle= \left(15~\ln N\right)^\alpha$ with $\alpha=1$ in (a) and $\langle n \rangle= \left(15~\ln N\right)^\alpha$ with $\alpha=2$ in (b) respectively. The large colored dots indicate their intersection. Here $L(x)$ and $D(x+y)$ are scaled by $\ln N$ for ease of comparison. Labelled are  the curves $y_{\rm c1}$ ($y_{\rm c2}$) which correspond to the largest (smallest) value of $y$ indicating a stable $L(x)=D(x+y)$ solution in the subcritical regime. Further $y_{\rm c3}$ gives the point where the networks breaks completely apart to become structurally disconnected meaning there is no giant set of nodes that can connect to each other with any fidelity. 
We generated one \er  network for each value of $N$, then $D(x+y)$ was determined by removing each link of the network with probability $1-p=1-e^{-(x+y)}$. $D(x+y)$ was computed based on 100 runs for each value of $x+y$. The intersection between the two nodes at $x_0+y$, is marked by blue dots for solutions in the supercritcal regime, and red dots for solutions in the subcritical regime. Next the size of the {\it backbone} (number of nodes in the backbone), $S(x_0+y)$, is plotted as a function of $y$ in (c,d) for $\alpha=1,2$ respectively. The functionally connected regime is represented as the blue region while the functionally subcritical  regime is shown as the light red region. The blue/red stripped area represents the region where both the functionally connected and subcritical regimes are stable. The dark red region on the other hand represents the structurally disconnected regime for a network of size $N=10^5$. 
}}
\label{fig3}
\end{figure*}

 \begin{figure*}

\centering
\includegraphics[width=1\textwidth]{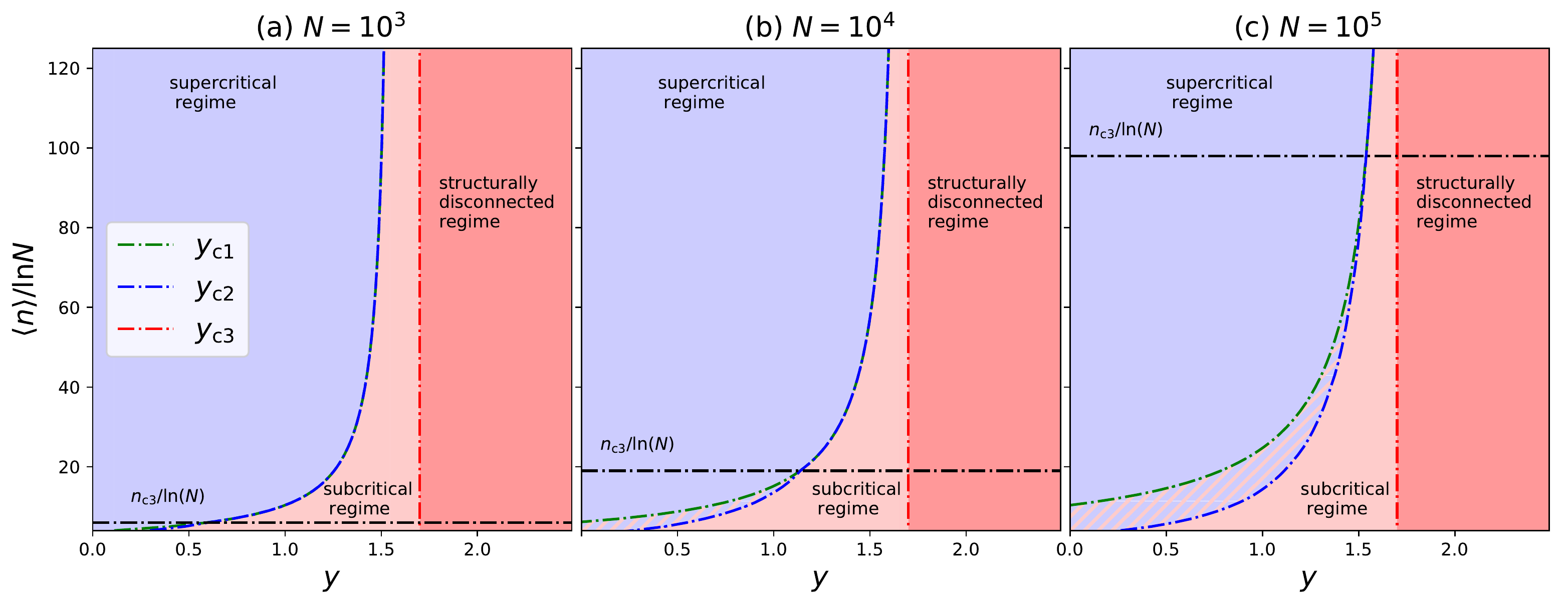}

\caption{{{\bf Phase diagram of a quantum \er  network} --- Depicted are phase diagrams of $\langle n \rangle / \ln N$ versus $y$ for a quantum \er network network comprised of $N = 10^4$ (a), $10^5$ (b) and $10^6$ (c) nodes respectively with  $c=6$ average degree and $\alpha=1$. Here we assume the number of entangled pairs in each link follows an exponential distribution with mean $\langle n \rangle$. The functionally supercritcal regime is represented as the blue region while the functionally subcritical regime is the light red region. Further the stripped (blue / light red) area represents the region where both the functionally connected and subcritical regime are stable. The dark red region represents the structurally disconnected regime. Next $y_{\rm c3}$ is the point where the networks breaks completely apart, meaning there is no giant set of nodes that can connect to each other with any fidelity. Also shown are the critical mean resource number $n_{c3} / \ln N$ for which the discontinuous phase transition is avoided.
}
} \label{fig4}
\end{figure*}

There are of course many well known network models we could examine with the \er  model network probably being the simplest~\cite{erdds1959random,newman2018networks,barabasi2016network,dorogovtsev2008critical} . This model has been well explored in complex network theory and  as such it is a good starting point, especially as one can compute $L(x)$ and $D(x+y)$ analytically when the number of nodes $N\rightarrow \infty$ (asymptotic limit). Considering only bond percolation (loss of links) we show in Fig.\ \ref{fig3} that the quantum backbone for a large \er   network is prone to an abrupt phase transition. Remembering that our parameter $y$ is associated with the logarithm of the success probability that no random link failures have occurred (given sufficient resources), we observe that the size of the {\it quantum backbone} actually drops abruptly as random link failures $y$ increase between two critical values $y_{c1}<y<y_{c2}$. Here $y_{\rm c1}, (y_{\rm c2}$) corresponds to the largest, (smallest) value of satisfying $L(x)=D(x+y)$. In this region the supercritical  (large {\it quantum backbone}) and subcritical (small {\it quantum backbone}) regimes are both stable solutions. This corresponds to a hysteresis region  whose span grows with the size of the network as shown in Fig.~\ref{fig4} (see Appendix~\ref{appendix2} for the analytical calculations). We observe that when the average number of entangled pairs in each link of the network is larger than a critical value $n_{\rm c}$ the traditional percolation phase transition is recovered as shown in Fig.~\ref{fig4} (see Appendix \ref{appendix2}). The value of $n_{\rm c}$ increases quickly with the size of the network (see Appendix~\ref{appendix3}).  It is useful to mention that we have used $\alpha=1$ as a conservative value.  As $\alpha$ increases the network in with the discontinuous phase transition is observed will be smaller. 

\begin{figure*}
\centering
\includegraphics[width=1\textwidth]{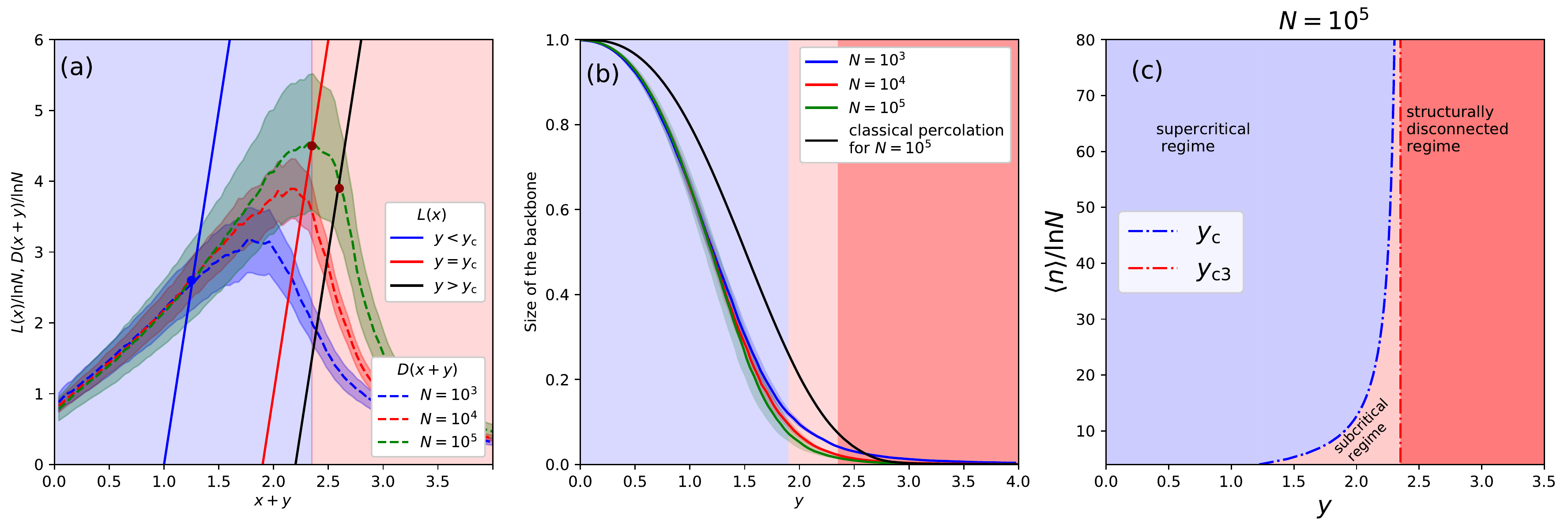}
\caption{{{\bf Robustness of a quantum \ba network} --- Exploration of  bond percolation on a quantum \ba  network with $N$ nodes and average degree  $c=6$ where the number of entangled pairs in each link follow an exponential distribution with mean $\langle n \rangle$. {We simulated the  the  Bollob\'as variant of the \ba model~\cite{Bollobas2004}} where we generated one \ba network for each value of $N$. $D(x+y)$ was then determined by removing link from the network with probability $1-p=1-e^{-(x+y)}$, based on 100 runs for each value of $x+y$.  We plot $L(x)$ and $D(x+y)$ versus $x+y$ for $\langle n \rangle= 60^\alpha$ with $\alpha=1$ in (a) with the large colored dots indicate their intersection. Labelled are  the curves $y_{\rm c}$ which correspond to the largest value of $y$ indicating a stable $L(x)=D(x+y)$ solution. The intersection between two nodes at $x_0+y$, is marked by blue dots for solutions in the supercritical regime, and red dots in the subcritical regime. Next in (b) we plot the size of the {\it backbone} (number of nodes in the backbone) as a function of $y$. The functionally supercritical regime is represented as the blue region while the functionally subcritical regime is shown as the light red region. The dark red region on the other hand represents the structural subcritical region for a network of size $N=10^5$. We immediately observe that no discontinuous phase transition is seen (unlike in the \er  case). Finally in (c) we depict the phase diagrams of $\langle n \rangle / \ln N$ versus $y$ for $N = 10^5$ nodes including the curves for $y_c$ and $y_{c3}$.}
}
\label{fig6}
\end{figure*}

\section{Quantum Scale-Free Networks}
These observations lead to a natural question about how general our results are -- especially in terms of the network model. As such it is useful explore the scale-free \ba quantum network whose classical counterpart is known to be more robust than the \er \cite{newman2018networks,barabasi2016network,dorogovtsev2008critical,Albert2000,PhysRevLett.85.5468}. In Fig.~\ref{fig6}(a) we plot $L(x)$ and $D(x+y)$ versus $x+y$ for various $N$ with $\mean{n}=30\ln(N)$ and $\alpha=1$ where the $\ln(N)$ scaling for $\mean{n}$ is chosen to make the comparison with the \er network easier. Our results for the \ba network show that we are in the supercritcal regime for much larger values of $x+y$ which highlights it robustness to errors. This is to be expected, the \ba network, and other scale-free networks, are known to be more robust than an \er network \cite{PhysRevLett.85.5468,Barabasi1999}. More importantly in Fig.~\ref{fig6} we observe no discontinuous phase transition in the  $N=10^3$ to $10^5$ region (unlike what occurred in the \er  situation Fig.~\ref{fig3}(c,d). This is exemplified in Fig.~\ref{fig6}(c) by the absence of a region where both the subcritical and supercritical are stable solutions of (\ref{eqdiamter_3}). In fact, one can show that there is always a value $\alpha_c$ such that for $\alpha < \alpha_c$ the discontinuous phase transition is suppressed (our network used in Fig.~\ref{fig6} has $\alpha_c>1$). It is useful to explore this $\alpha_c$ parameter in a little more details.  When our resources are exponentially distributed, it is straightforward to show (see Appendix \ref{appendix4}) that $\alpha_c$ is given by
\begin{equation} 
\alpha_c=\min_{x<y_{c3}}  \frac{D(x)/x}{dD(x)/dx}
\end{equation}
which establishes the existence of a sufficient repeater efficiency so that most of the classical behavior is recovered. Despite the suppression of the discontinuous phase transition for $\alpha < \alpha_c$ there are still a few differences between the various quantum cases. Unlike what one expects for a typical  \ba network ~\cite{PhysRevLett.85.5468,dorogovtsev2008critical,barabasi2016network} the point at each the networks breaks apart, $y_c$, doesn't change significantly with the network size. To understand this, it is useful to look at the relation between $y_c$ and $y_{c3}={-\ln}(p_c)$ with $p_c$ being the percolation threshold of the network (which for the usual \ba network tends to zero as $N$ increases).  When our resources are exponentially distributed $y_c$ (or $y_{c1}$ for a discontinuous phase transition) is related to $y_{c3}$ via, (see Appendix \ref{appendix3})
 \begin{equation} 
 y_{c(c_1)}=y_{c3} -\frac{\left(D_{\rm max}\right)^\alpha}{\mean{n}}.
\label{eqdiamter_4} 
\end{equation}
This provides quite an interesting insight to to this apparent change of behavior. It is well known that $y_{c3}$  grows with the network size for the \ba network \cite{PhysRevLett.85.5468,dorogovtsev2008critical,barabasi2016network}, but so does $D_{\rm max}$ (this is what prevents the suppression of the phase-transition).  This means that increase of $y_c$ can be mitigated by increasing $\mean{n}$ proportionally with $(D_{\rm max})^{\alpha}$.

\section{Network metrics}

Our exploration of the \er  and scale-free Bollob\'as quantum networks has highlighted how the topology of these networks plays a significant role in its robustness but how? We need to quantify this behaviour using three important characterization parameters:
 \begin{itemize}
 \item First is $y_{c3}$ which is the number of links needed to be removed before the network breaks as $\mean{n}\rightarrow \infty$. 
 \item Second is $D_{\rm max} \equiv \max_{x+y>0} D^{\alpha}(x+y)$ which measures the degree to which finite resources in the network changes $y_{ c(c_1)}$. 
 \item Third is $\alpha_c$ which measures how efficient a repeater protocol needs to be in order to  avoid the discontinuous phase transition for a given network topology. 
 \end{itemize}

\begin{figure*}
\centering
\includegraphics[width=1.\textwidth]{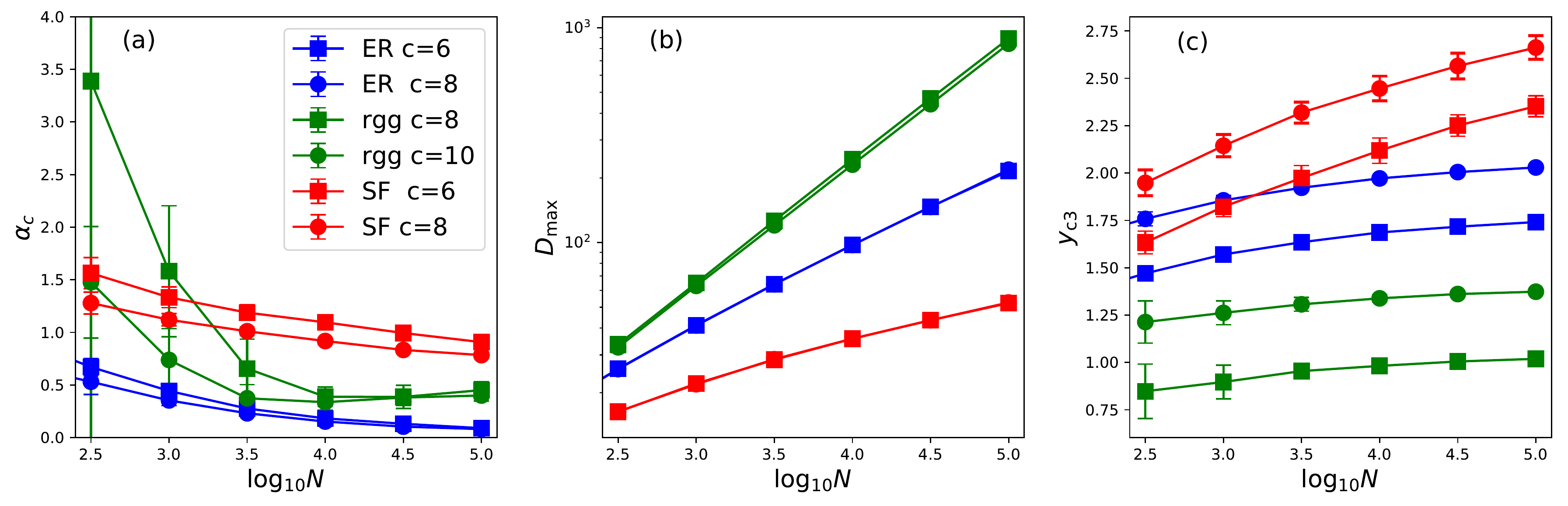}
\caption{{\textbf{Robustness of the network as a function of its topology} --- Shown are the three critical parameters $\alpha_c$ (a),  $D_{\rm max}$ (b)  and $y_{\rm c3}$ (c) used to determine the robustness are shown for the \er  (ER), \ba  (SF) and geometric graphs (rgg) networks for N varying between $10^{2.5} -  10^5$.}  For a network to be robust we want $\alpha_c$ and $y_{\rm c3}$ to be as large as possible with $D_{\rm max}$ to be as small as possible.  We estimate the parameters above based on 100 network realization for each network model and value of $c$. For each network $D(x)$ was computed based on 100 runs. \label{fig5}
}
\end{figure*}

The first two parameters are related to how many links need to be removed before the network breaks apart while the third is associated with the efficiency of the repeater protocol. These three parameters can be determined for  both the \er  network and \ba networks. It is also useful to determine them for another important case of networks associated  with geometric random graphs as well~\cite{PhysRevE.66.016121}.   Long direct quantum links are extremely costly to create, with a success probability that decreases with the distance between stations. 
Although one can have a link between two station that is composed by smaller elementary links, such a solution would be costly, specially for large distances. One of the key issues with a quantum \ba network is the presence of long range links which would be  costly to create.  Entanglement has been directly distributed over 300 km links \cite{sasaki2011field} but the rate exponentially decreases with distance. Although one can have a link between two station that is composed by smaller elementary links, such a solution would be costly, specially for large distances. Geometric random networks were designed to handle exactly this type of situations.  It is a spacial network that only contains links between nodes that are close to each other\cite{PhysRevE.66.016121}, making it a very natural model for quantum networks. With these three network topologies in mind ---  Erd\H{o}s-R\'enyi, Barab{\'a}si-Albert, and geometric random graphs --- we plot in Fig.~\ref{fig5} the quantities $y_{c3}$,  $D_{\rm max}$, and $\alpha_c$ versus $N$, for values of average degree $c=6, 8$.  Our plots clearly show that  scale-free networks are more robust according to all three parameters, and is the only network that for the selected parameters is able to avoid the discontinuous phase transition for $\alpha>1$ and $N>10^3$. Further the \ba network was the best performing in terms of all three parameters for larger networks sizes ($N \geq 10^{3.5}$). These results show the importance of choosing the right network topology.

There is one more important consideration we must address here in terms of the generality of our results. This relates to our choice of an exponential resource distribution we used throughout the paper. This was primarily chosen for the ease of our calculations. Other distributions, Poisson and Guassian for instance, show similar network behavior and our conclusion about the robustness of the \ba networks remains unchanged (see Appendix \ref{appendix3}).
The form of the quantum repeaters, their operation and how engineers of the future quantum Internet distribute resources through the network will determine what the resource distribution actually is. It is still an open question as to what the optimal resource distribution actually would be.


\section{Conclusions}

Quantum networks are a new paradigm of complex networks whose properties are governed by the laws of quantum physics. The traditional Bernoulli percolation-based tools used in classical networks cannot be used to study the behaviour and robustness of quantum networks. By introducing a new concept, the quantum backbone, we derived a metric to measure the connectivity of a quantum network, and have shown how large-scale quantum networks based on noisy quantum-repeater nodes connected by noisy channels are prone to discontinuous phase transitions. This abrupt behaviour breaks the network into disconnected pieces, severely limiting its operational reach. However, we have established how the right network topology, combined with advanced quantum repeater architectures \cite{Jiang2009,munro2012,Muralidharan2014}, can contribute to this robustness. Our results provide guiding criteria for the design and development of a robust large-scale quantum Internet.

{\bf{Acknowledgements:}} The authors thank Akshat Kumar for his feedback on appendix \ref{appendix1} and acknowledge the support from the JTF project \textit{The Nature of Quantum Networks} (ID 60478).~Furthermore, BC and YO thank the support from Funda\c{c}\~{a}o para a Ci\^{e}ncia e a Tecnologia (Portugal), namely through projects UIDB/EEA/50008/2020 and QuantSat-PT, as well as from projects TheBlinQC and QuantHEP supported by the EU H2020 QuantERA ERA-NET Cofund in Quantum Technologies and by FCT (QuantERA/0001/2017 and QuantERA/0001/2019, respectively), and from the EU H2020 Quantum Flagship projects QIA (820445) and QMiCS (820505).

\appendix

\section{Relation between functionally connected component and the  maximum clique}
\label{appendix1}

If a problem is at least as hard as an NP-complete, then the problem is NP-hard (and can or not be NP-complete). If for all pairs of ${ij}$, $n_{ij}$ is either  0 or 1,  then our problem is equivalent to solving the maximum clique problem that is a known NP-complete problem \cite{karp1972reducibility}. Therefore we can conclude that finding the largest functionally connected component is at least as hard as solving the maximum clique problem making this problem NP-hard.

\section{Analytical calculations for an \er network}
\label{appendix3}
In this appendix we will calculate $L(x)$ and $D(x+y)$ analytically for an $\rightarrow \gg 1$ node network with exponential distribution of entangled qubits across the links. Let us consider an \er network with $N$ where links between two nodes exists with probability $c/(N-1)$, corresponding to an average degree $c$.  The state of the art approximation for the diameter of an \er network  can be found in~\cite{riordan2010diameter}. For the supercritical regime ($c>1$),  is given by
\begin{equation}
d/\ln N=\frac{1}{\ln c}-\frac{2}{\ln~c^*}+\frac{O(1)}{\ln N}
\end{equation}
where $c^*$ is the solution between $0<c^*\leq 1$ for $c^*\exp(-c^*)=c\exp(-c)$. For $\ln N \gg 1$ the last term can be ignored. 
Next in the subcritical regime $c<1$, far enough from the critical point,
\begin{align}
\frac{d}{\ln N}=-\frac{1}{\ln c}-\frac{2}{\ln c \ln N}-\frac{O(1)}{\ln N}
\end{align}
For $\ln N\gg 1$ (and therefore $N\gg 1$) we arrive at the following equation,
\begin{equation}
d/\ln N=
\begin{cases} 
\frac{1}{\ln c}-\frac{2}{\ln~c^*}& c>1\\
-\frac{1}{\ln c}& c<1.
\end{cases}
\label{ER_LHD}
\end{equation}
Where we note  that this is scaling is more rigorous than the traditional scaling $\ln N/\ln c$, found in most books~\cite{barabasi2016network}. When comparing with simulation we found the extra term to be not negligible. 

\begin{figure}[t!]
\centering
\includegraphics[width=0.44\textwidth]{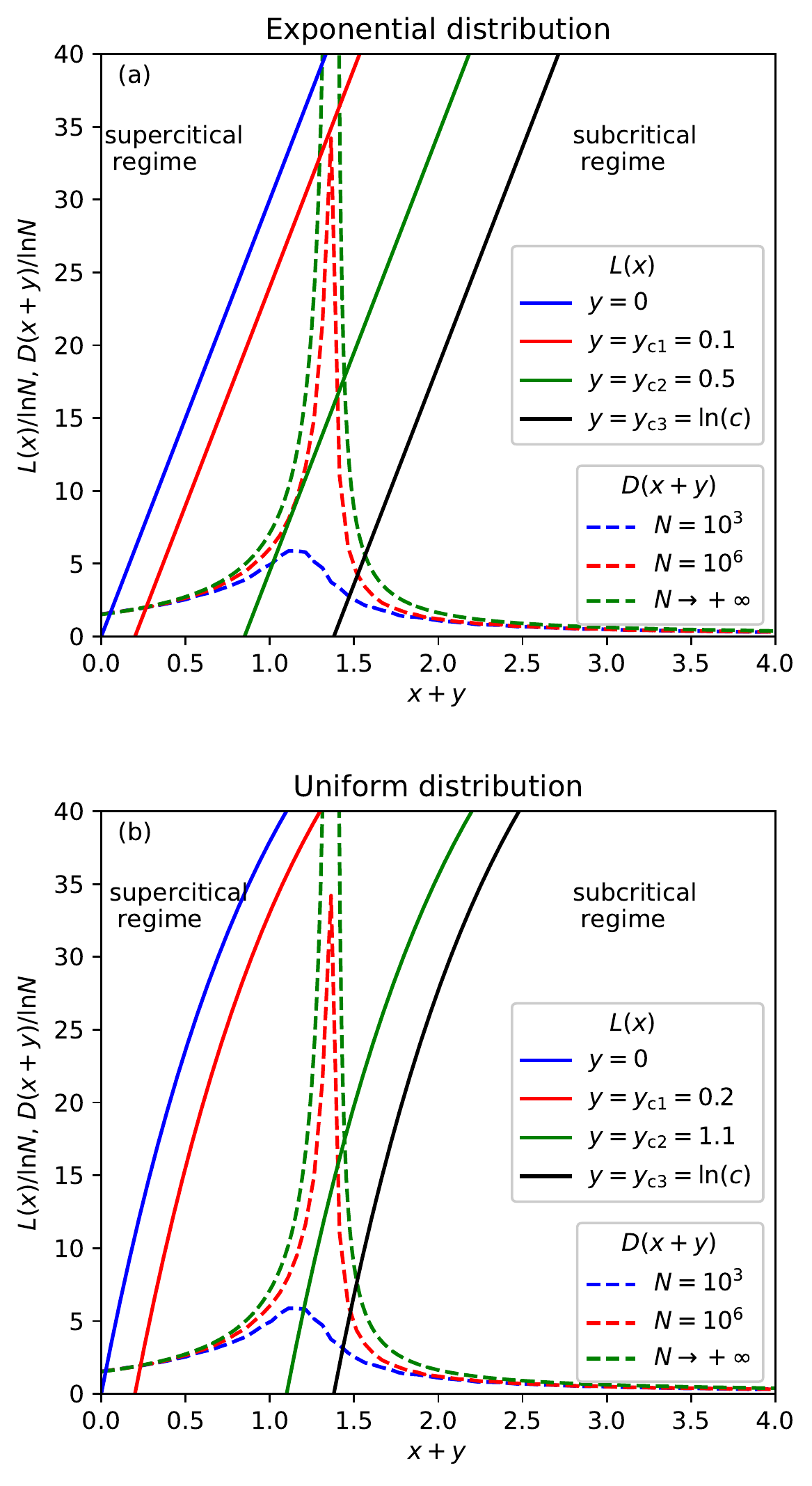}
\caption{{\bf Asymptomatic robustness of a quantum Erd\H{o}s-R\'enyi network}. Exploration of an $N$ nodes bond percolation for a \er network with average degree  $c=4$ where the number of entangled pairs in each edge follow an exponential distribution with mean number $\langle n \rangle$ in (a) and here the number of entangled pairs in each edge follow an uniform distribution with mean number $\langle n \rangle$ in (b). We plot $L(x)$ and $D(x+y)$ versus $x+y$ for $\langle n \rangle= \left(30 \ln N\right)^\alpha$ with $\alpha=1$. Theoretical prediction for $D(x+y)$ when $N\rightarrow +\infty$ shown as a green dashed curve.\label{fig3_SI} }
\end{figure}

Let us now consider an example of an  \er network, with $N$ nodes and an average degree $c$. Considering that links remain in the network with  $p=e^{-(x+y)}$ which is the probability that we have sufficient pairs to create our link (given by $p_x=e^{-x}$) and that their have not been any random failures of those links ($p_y=e^{-y}$)  we can establish that $D(x+y)$ is given by,
 \begin{equation}
\frac{D(x+y)}{\ln N}=
\begin{cases} 
\frac{1}{\ln c-(x+y)}-\frac{2}{\ln~c^*}& c~e^{-(x+y)}>1\\
-\frac{1}{\ln c-(x+y)}& c~e^{-(x+y)}<1,
\end{cases}
\label{ER_diameter_2}
\end{equation}
where $c^*$ is the solution of $c^*\exp(-c^*)=ce^{-(x+y)}\exp(-ce^{-(x+y)})$ in the range $0<c^*\leq 1$.
For simplicity let us consider and  the number entangled qubit pairs in each edge follows an exponential probability distribution , with an average of $\langle n \rangle$ of entangled qubits in each edge (which is our distribution used in the main text). The probability that an edge contain more than $l_{\rm fixed}^\alpha$ entangled qubit pairs is given by $p(l_{\rm fixed}^\alpha)=\exp\left(-l_{\rm fixed}^\alpha/\langle n \rangle\right)$. In turn this means, 

\begin{equation}
L(x)=x^{1/\alpha} \mean{n}^{1/\alpha},
\label{ER_RHD}
\end{equation}
where $L(x)$ which is the distance $l_{\rm fixed}$ written as a function of $x\equiv-\ln(p_{\rm fixed})$. Combining Eq.\ (\ref{ER_RHD}) with with Eq.\ (\ref{ER_diameter_2}), we obtain
\begin{equation}
x_0^{1/\alpha} \frac{\mean{n}^{1/\alpha}}{\ln N}=\begin{cases} 
\frac{\ln N}{\ln c-(x_0+y)}-2 \frac{\ln N}{\ln~c^*}& c~e^{-\tilde{x}_0}>1\\
-\frac{\ln N}{\ln c-(x_0+y)}& c~e^{-\tilde{x_0}}<1.
\label{ER_diameter_3}
\end{cases}
\end{equation}
From here we can derive that to maintain the proprieties of the network constant, for large number of nodes $N$, the average number of qubits  should grow with the number nodes as $\mean{n}\sim (\ln N)^\alpha$. Solving  Eq.\ (\ref{ER_diameter_3}) yields the same behavior displayed in Fig.~\ref{fig2}, with the difference that the solution in the subcritical regime is always present, as shown in Fig.~\ref{fig3_SI}.

If we consider a different distribution of entangled qubits the results will change qualitatively but not quantitatively. Let us  consider an uniform distribution of entangled qubits across the. In such case 
\begin{equation} 
L(x)=\left[2 \mean{n}\left(e^{-x}-1\right)\right]^{1/\alpha}.
\end{equation} 
Combining Eq.\ (\ref{ER_RHD}) with with Eq.\ (\ref{ER_diameter_3}), one obtains,

\begin{widetext}
\begin{equation}
\left(e^{-x_0}-1\right)^{1/\alpha} \frac{(2\mean{n})^{1/\alpha}}{\ln N}=\begin{cases} 
\frac{\ln N}{\ln c-(x_0+y)}-2 \frac{\ln N}{\ln~c^*} c e^{-(x_0+y)}>1
-\frac{\ln N}{\ln c-\tilde{x}_0}& ce^{-(x_0+y)}<1.
\end{cases}
\label{master_eq_udn}
\end{equation}
\end{widetext}

This gives the same resource scaling we have previously seen. Further $L(x)$ now have a polynomial scaling as seen in Fig.~\ref{fig3_SI}.

\section{Conditions for the suppression of the first order phase-transition}
\label{appendix2}
In general, for an exponential distribution of entangled qubit pairs, Eq.\ (\ref{eqdiamter_3}) can be written as,
\begin{equation}
(\tilde{x}-y)\mean{n}=D(\tilde{x})^{\alpha}
\label{chapter5::eq1}
\end{equation} 
where we define $\tilde{x}=x+y$. If for each value of $y$ there is only solution for Eq.\ (\ref{chapter5::eq1}) then this phase-transition will be continuous. To check if the phase transition is continuous one  can write y as a function of $\tilde{x}$ as
\begin{equation}
Y(\tilde{x})=\tilde{x}-\frac{D(\tilde{x})^{\alpha}}{\mean{n}}
\label{y_eq}
\end{equation} 
Given a value of $y$, $\tilde{x}_0\equiv x_0+y$ can be found from the equation $Y(\tilde{x}_0)=y$. If $Y(\tilde{x})$ is a monotonically decreasing function for $0<\tilde{x}<y_{c_3}$, then the transition is continuous. The region for $\tilde{x}\geq y_{c_3}$ is not important since at this point the network is disconnected. In contrast, if the function is not a monotonically decreasing function, then one can say that in general the phase transition will not be continuous. For $\alpha\neq 0$ the phase transition is only continuous if 
\begin{equation}
\mean{n}\geq n_{\rm c},
\end{equation} 
where, 
\begin{equation}
n_{\rm c}=\alpha\max_{0<\tilde{x}<y_{c_3}}\left[D(\tilde{x})^{\alpha-1} \frac{d}{d\tilde{x}} D(\tilde{x})\right].
\label{eq:critical_resources}
\end{equation} 
Furthermore, there is a value $\alpha_c$  such that for  $\alpha<\alpha_c$, there will be no discontinuous phase transition for any value of $\mean{n}$. To derive such a value one needs to consider that that network is in the supercritcal regime when no links are removed from  the network. Given a value of $\tilde{x}<y_{c_3}$ on can compute the smaller value of average resources $n^*$ that guarantees the network is in the supercritcal regime regime for $y=0$, given by 
\begin{equation}
\mean{n}^{\*}=D(\tilde{x})^{\alpha}/\tilde{x}.
\end{equation} 
Now, from Eq.\ (\ref{eq:critical_resources}), one can consider $\alpha_c$ as the value of $\alpha$ when $n_{\rm c}=\mean{n}^*$, to obtain, 

\begin{equation}
\alpha_c=\min_{x<y_{c_3}}  \frac{D(x)/x}{dD(x)/dx}
\end{equation} 

\section{Relation between $y_{c3}$ and $y_{c1}$}
\label{appendix4}
$y_{\rm c1}$ gives the point where the networks breaks functionally, and $y_{\rm c3}$ gives the point the network is not in the functionally supercritcal regime.  Considering that our resources are exponentially distributed, the relation between $y_{\rm c1}$ and $y_{\rm c3}$ can be computed using Eq.~\ref{chapter5::eq1}, and is given by
\begin{equation}
y_{\rm c1}=Y(y_{\rm c3})=y_{\rm c3}-\frac{D(y_{\rm c3})^{\alpha}}{\mean{n}}
\end{equation} 
Using the fact the diameter of the network diverges at $y_{\rm c3}$ $D_{\rm max}=D(y_{\rm c3})$, one obtains Eq.~(\ref{eqdiamter_4}).


\begin{thebibliography}{58}
\expandafter\ifx\csname natexlab\endcsname\relax\def\natexlab#1{#1}\fi
\expandafter\ifx\csname bibnamefont\endcsname\relax
  \def\bibnamefont#1{#1}\fi
\expandafter\ifx\csname bibfnamefont\endcsname\relax
  \def\bibfnamefont#1{#1}\fi
\expandafter\ifx\csname citenamefont\endcsname\relax
  \def\citenamefont#1{#1}\fi
\expandafter\ifx\csname url\endcsname\relax
  \def\url#1{\texttt{#1}}\fi
\expandafter\ifx\csname urlprefix\endcsname\relax\def\urlprefix{URL }\fi
\providecommand{\bibinfo}[2]{#2}
\providecommand{\eprint}[2][]{\url{#2}}

\bibitem[{\citenamefont{Kimble}(2008)}]{Kimble2008}
\bibinfo{author}{\bibfnamefont{H.~J.} \bibnamefont{Kimble}},
\bibinfo{journal}{Nature} \textbf{\bibinfo{volume}{453}},
\bibinfo{pages}{1023} (\bibinfo{year}{2008}).

\bibitem[{\citenamefont{Caleffi et~al.}(2018)\citenamefont{Caleffi,
  Cacciapuoti, and Bianchi}}]{Caleffi2018}
\bibinfo{author}{\bibfnamefont{M.}~\bibnamefont{Caleffi}},
\bibinfo{author}{\bibfnamefont{A.~S.} \bibnamefont{Cacciapuoti}},
\bibnamefont{and} \bibinfo{author}{\bibfnamefont{G.}~\bibnamefont{Bianchi}}
 (\bibinfo{year}{2018}).

\bibitem[{\citenamefont{Van~Meter}(2014)}]{vanmeter2014}
\bibinfo{author}{\bibfnamefont{R.}~\bibnamefont{Van~Meter}},
  \emph{\bibinfo{title}{Quantum Networking}} (\bibinfo{publisher}{Hoboken:
  Wiley}, \bibinfo{year}{2014}).

\bibitem[{\citenamefont{Perseguers et~al.}(2010)\citenamefont{Perseguers,
  Lewenstein, Acin, and Cirac}}]{citeulike:7231538}
\bibinfo{author}{\bibfnamefont{S.}~\bibnamefont{Perseguers}},
  \bibinfo{author}{\bibfnamefont{M.}~\bibnamefont{Lewenstein}},
  \bibinfo{author}{\bibfnamefont{A.}~\bibnamefont{Acin}}, \bibnamefont{and}
  \bibinfo{author}{\bibfnamefont{J.~I.} \bibnamefont{Cirac}},
  \bibinfo{journal}{Nat Phys} \textbf{\bibinfo{volume}{6}},
  \bibinfo{pages}{539} (\bibinfo{year}{2010}), ISSN \bibinfo{issn}{1745-2473}.

\bibitem[{\citenamefont{Inlek et~al.}(2017)\citenamefont{Inlek, Crocker,
  Lichtman, Sosnova, and Monroe}}]{inlek2017multispecies}
\bibinfo{author}{\bibfnamefont{I.~V.} \bibnamefont{Inlek}},
  \bibinfo{author}{\bibfnamefont{C.}~\bibnamefont{Crocker}},
  \bibinfo{author}{\bibfnamefont{M.}~\bibnamefont{Lichtman}},
  \bibinfo{author}{\bibfnamefont{K.}~\bibnamefont{Sosnova}}, \bibnamefont{and}
  \bibinfo{author}{\bibfnamefont{C.}~\bibnamefont{Monroe}},
  \bibinfo{journal}{Phys. Rev. Lett.} \textbf{\bibinfo{volume}{118}},
  \bibinfo{pages}{250502} (\bibinfo{year}{2017}).

\bibitem[{\citenamefont{Briegel et~al.}(1998)\citenamefont{Briegel, D\"ur,
  Cirac, and Zoller}}]{PhysRevLett.81.5932}
\bibinfo{author}{\bibfnamefont{H.-J.} \bibnamefont{Briegel}},
  \bibinfo{author}{\bibfnamefont{W.}~\bibnamefont{D\"ur}},
  \bibinfo{author}{\bibfnamefont{J.~I.} \bibnamefont{Cirac}}, \bibnamefont{and}
  \bibinfo{author}{\bibfnamefont{P.}~\bibnamefont{Zoller}},
  \bibinfo{journal}{Phys. Rev. Lett.} \textbf{\bibinfo{volume}{81}},
  \bibinfo{pages}{5932} (\bibinfo{year}{1998}).

\bibitem[{\citenamefont{Brukner}(2014)}]{brukner2014quantum}
\bibinfo{author}{\bibfnamefont{{\v{C}}.}~\bibnamefont{Brukner}},
  \bibinfo{journal}{Nat. Physics} \textbf{\bibinfo{volume}{10}},
  \bibinfo{pages}{259} (\bibinfo{year}{2014}).

\bibitem[{\citenamefont{Chakraborty et~al.}(2017)\citenamefont{Chakraborty,
  Novo, Di~Giorgio, and Omar}}]{PhysRevLett.119.220503}
\bibinfo{author}{\bibfnamefont{S.}~\bibnamefont{Chakraborty}},
  \bibinfo{author}{\bibfnamefont{L.}~\bibnamefont{Novo}},
  \bibinfo{author}{\bibfnamefont{S.}~\bibnamefont{Di~Giorgio}},
  \bibnamefont{and} \bibinfo{author}{\bibfnamefont{Y.}~\bibnamefont{Omar}},
  \bibinfo{journal}{Phys. Rev. Lett.} \textbf{\bibinfo{volume}{119}},
  \bibinfo{pages}{220503} (\bibinfo{year}{2017}).

\bibitem[{\citenamefont{Brito et~al.}(2020)\citenamefont{Brito, Canabarro,
  Chaves, and Cavalcanti}}]{PhysRevLett.124.210501}
\bibinfo{author}{\bibfnamefont{S.}~\bibnamefont{Brito}},
  \bibinfo{author}{\bibfnamefont{A.}~\bibnamefont{Canabarro}},
  \bibinfo{author}{\bibfnamefont{R.}~\bibnamefont{Chaves}}, \bibnamefont{and}
  \bibinfo{author}{\bibfnamefont{D.}~\bibnamefont{Cavalcanti}},
  \bibinfo{journal}{Phys. Rev. Lett.} \textbf{\bibinfo{volume}{124}},
  \bibinfo{pages}{210501} (\bibinfo{year}{2020}).

\bibitem[{\citenamefont{Pirandola}(2019{\natexlab{a}})}]{Pirandola2019_1}
\bibinfo{author}{\bibfnamefont{S.}~\bibnamefont{Pirandola}},
  \bibinfo{journal}{Communications Physics} \textbf{\bibinfo{volume}{2}}
  (\bibinfo{year}{2019}{\natexlab{a}}), ISSN \bibinfo{issn}{23993650}.

\bibitem[{\citenamefont{Pirandola}(2019{\natexlab{b}})}]{Pirandola2019_2}
\bibinfo{author}{\bibfnamefont{S.}~\bibnamefont{Pirandola}},
  \bibinfo{journal}{arXiv}  (\bibinfo{year}{2019}{\natexlab{b}}),
  \eprint{1912.11355}.

\bibitem[{\citenamefont{Wu et~al.}(2020)\citenamefont{Wu, Tian, Coutinho, Omar,
  and Liu}}]{PhysRevA.101.052315}
\bibinfo{author}{\bibfnamefont{A.-K.} \bibnamefont{Wu}},
  \bibinfo{author}{\bibfnamefont{L.}~\bibnamefont{Tian}},
  \bibinfo{author}{\bibfnamefont{B.~C.} \bibnamefont{Coutinho}},
  \bibinfo{author}{\bibfnamefont{Y.}~\bibnamefont{Omar}}, \bibnamefont{and}
  \bibinfo{author}{\bibfnamefont{Y.-Y.} \bibnamefont{Liu}},
  \bibinfo{journal}{Phys. Rev. A} \textbf{\bibinfo{volume}{101}},
  \bibinfo{pages}{052315} (\bibinfo{year}{2020}).

\bibitem[{\citenamefont{Chakraborty et~al.}(2016)\citenamefont{Chakraborty,
  Novo, Ambainis, and Omar}}]{PhysRevLett.116.100501}
\bibinfo{author}{\bibfnamefont{S.}~\bibnamefont{Chakraborty}},
  \bibinfo{author}{\bibfnamefont{L.}~\bibnamefont{Novo}},
  \bibinfo{author}{\bibfnamefont{A.}~\bibnamefont{Ambainis}}, \bibnamefont{and}
  \bibinfo{author}{\bibfnamefont{Y.}~\bibnamefont{Omar}},
  \bibinfo{journal}{Phys. Rev. Lett.} \textbf{\bibinfo{volume}{116}},
  \bibinfo{pages}{100501} (\bibinfo{year}{2016}).

\bibitem[{\citenamefont{Faccin et~al.}(2014)\citenamefont{Faccin, Migda\l{},
  Johnson, Bergholm, and Biamonte}}]{PhysRevX.4.041012}
\bibinfo{author}{\bibfnamefont{M.}~\bibnamefont{Faccin}},
  \bibinfo{author}{\bibfnamefont{P.}~\bibnamefont{Migda\l{}}},
  \bibinfo{author}{\bibfnamefont{T.~H.} \bibnamefont{Johnson}},
  \bibinfo{author}{\bibfnamefont{V.}~\bibnamefont{Bergholm}}, \bibnamefont{and}
  \bibinfo{author}{\bibfnamefont{J.~D.} \bibnamefont{Biamonte}},
  \bibinfo{journal}{Phys. Rev. X} \textbf{\bibinfo{volume}{4}},
  \bibinfo{pages}{041012} (\bibinfo{year}{2014}).

\bibitem[{\citenamefont{Biamonte et~al.}(2019)\citenamefont{Biamonte, Faccin,
  and {De Domenico}}}]{Biamonte2019}
\bibinfo{author}{\bibfnamefont{J.}~\bibnamefont{Biamonte}},
  \bibinfo{author}{\bibfnamefont{M.}~\bibnamefont{Faccin}}, \bibnamefont{and}
  \bibinfo{author}{\bibfnamefont{M.}~\bibnamefont{{De Domenico}}},
  \bibinfo{journal}{Commun. Phys.} \textbf{\bibinfo{volume}{2}},
  \bibinfo{pages}{1} (\bibinfo{year}{2019}), ISSN \bibinfo{issn}{23993650},
  \eprint{1702.08459}.

\bibitem[{\citenamefont{Perseguers et~al.}(2008)\citenamefont{Perseguers,
  Cirac, Ac\'{\i}n, Lewenstein, and Wehr}}]{PhysRevA.77.022308}
\bibinfo{author}{\bibfnamefont{S.}~\bibnamefont{Perseguers}},
  \bibinfo{author}{\bibfnamefont{J.~I.} \bibnamefont{Cirac}},
  \bibinfo{author}{\bibfnamefont{A.}~\bibnamefont{Ac\'{\i}n}},
  \bibinfo{author}{\bibfnamefont{M.}~\bibnamefont{Lewenstein}},
  \bibnamefont{and} \bibinfo{author}{\bibfnamefont{J.}~\bibnamefont{Wehr}},
  \bibinfo{journal}{Phys. Rev. A} \textbf{\bibinfo{volume}{77}},
  \bibinfo{pages}{022308} (\bibinfo{year}{2008}).

\bibitem[{\citenamefont{Cuquet and Calsamiglia}(2009)}]{PhysRevLett.103.240503}
\bibinfo{author}{\bibfnamefont{M.}~\bibnamefont{Cuquet}} \bibnamefont{and}
  \bibinfo{author}{\bibfnamefont{J.}~\bibnamefont{Calsamiglia}},
  \bibinfo{journal}{Phys. Rev. Lett.} \textbf{\bibinfo{volume}{103}},
  \bibinfo{pages}{240503} (\bibinfo{year}{2009}).

\bibitem[{\citenamefont{Cuquet and Calsamiglia}(2011)}]{PhysRevA.83.032319}
\bibinfo{author}{\bibfnamefont{M.}~\bibnamefont{Cuquet}} \bibnamefont{and}
  \bibinfo{author}{\bibfnamefont{J.}~\bibnamefont{Calsamiglia}},
  \bibinfo{journal}{Phys. Rev. A} \textbf{\bibinfo{volume}{83}},
  \bibinfo{pages}{032319} (\bibinfo{year}{2011}).

\bibitem[{\citenamefont{Brito et~al.}(2021)\citenamefont{Brito, Canabarro,
  Cavalcanti, and Chaves}}]{PRXQuantum.2.010304}
\bibinfo{author}{\bibfnamefont{S.}~\bibnamefont{Brito}},
  \bibinfo{author}{\bibfnamefont{A.}~\bibnamefont{Canabarro}},
  \bibinfo{author}{\bibfnamefont{D.}~\bibnamefont{Cavalcanti}},
  \bibnamefont{and} \bibinfo{author}{\bibfnamefont{R.}~\bibnamefont{Chaves}},
  \bibinfo{journal}{PRX Quantum} \textbf{\bibinfo{volume}{2}},
  \bibinfo{pages}{010304} (\bibinfo{year}{2021}).

\bibitem[{\citenamefont{Gisin and Thew}(2007)}]{gisin2007quantum}
\bibinfo{author}{\bibfnamefont{N.}~\bibnamefont{Gisin}} \bibnamefont{and}
  \bibinfo{author}{\bibfnamefont{R.}~\bibnamefont{Thew}},
  \bibinfo{journal}{Nat. Photonics} \textbf{\bibinfo{volume}{1}},
  \bibinfo{pages}{165} (\bibinfo{year}{2007}).

\bibitem[{\citenamefont{Gisin et~al.}(2002)\citenamefont{Gisin, Ribordy,
  Tittel, and Zbinden}}]{Gisin2002}
\bibinfo{author}{\bibfnamefont{N.}~\bibnamefont{Gisin}},
  \bibinfo{author}{\bibfnamefont{G.}~\bibnamefont{Ribordy}},
  \bibinfo{author}{\bibfnamefont{W.}~\bibnamefont{Tittel}}, \bibnamefont{and}
  \bibinfo{author}{\bibfnamefont{H.}~\bibnamefont{Zbinden}},
  \bibinfo{journal}{Rev. Mod. Phys.} \textbf{\bibinfo{volume}{74}},
  \bibinfo{pages}{145} (\bibinfo{year}{2002}).

\bibitem[{\citenamefont{Bennett and DiVincenzo}(2000)}]{bennett2000quantum}
\bibinfo{author}{\bibfnamefont{C.~H.} \bibnamefont{Bennett}} \bibnamefont{and}
  \bibinfo{author}{\bibfnamefont{D.~P.} \bibnamefont{DiVincenzo}},
  \bibinfo{journal}{nature} \textbf{\bibinfo{volume}{404}},
  \bibinfo{pages}{247} (\bibinfo{year}{2000}).

\bibitem[{\citenamefont{Arute et~al.}(2019)\citenamefont{Arute, Arya, Babbush,
  Bacon, Bardin, Barends, Biswas, Boixo, Brandao, Buell
  et~al.}}]{arute2019quantum}
\bibinfo{author}{\bibfnamefont{F.}~\bibnamefont{Arute}},
  \bibinfo{author}{\bibfnamefont{K.}~\bibnamefont{Arya}},
  \bibinfo{author}{\bibfnamefont{R.}~\bibnamefont{Babbush}},
  \bibinfo{author}{\bibfnamefont{D.}~\bibnamefont{Bacon}},
  \bibinfo{author}{\bibfnamefont{J.~C.} \bibnamefont{Bardin}},
  \bibinfo{author}{\bibfnamefont{R.}~\bibnamefont{Barends}},
  \bibinfo{author}{\bibfnamefont{R.}~\bibnamefont{Biswas}},
  \bibinfo{author}{\bibfnamefont{S.}~\bibnamefont{Boixo}},
  \bibinfo{author}{\bibfnamefont{F.~G.} \bibnamefont{Brandao}},
  \bibinfo{author}{\bibfnamefont{D.~A.} \bibnamefont{Buell}},
  \bibnamefont{et~al.}, \bibinfo{journal}{Nature}
  \textbf{\bibinfo{volume}{574}}, \bibinfo{pages}{505} (\bibinfo{year}{2019}).

\bibitem[{\citenamefont{Zhong et~al.}(2020)\citenamefont{Zhong, Wang, Deng,
  Chen, Peng, Luo, Qin, Wu, Ding, Hu et~al.}}]{Zhong1460}
\bibinfo{author}{\bibfnamefont{H.-S.} \bibnamefont{Zhong}},
  \bibinfo{author}{\bibfnamefont{H.}~\bibnamefont{Wang}},
  \bibinfo{author}{\bibfnamefont{Y.-H.} \bibnamefont{Deng}},
  \bibinfo{author}{\bibfnamefont{M.-C.} \bibnamefont{Chen}},
  \bibinfo{author}{\bibfnamefont{L.-C.} \bibnamefont{Peng}},
  \bibinfo{author}{\bibfnamefont{Y.-H.} \bibnamefont{Luo}},
  \bibinfo{author}{\bibfnamefont{J.}~\bibnamefont{Qin}},
  \bibinfo{author}{\bibfnamefont{D.}~\bibnamefont{Wu}},
  \bibinfo{author}{\bibfnamefont{X.}~\bibnamefont{Ding}},
  \bibinfo{author}{\bibfnamefont{Y.}~\bibnamefont{Hu}}, \bibnamefont{et~al.},
  \bibinfo{journal}{Science} \textbf{\bibinfo{volume}{370}},
  \bibinfo{pages}{1460} (\bibinfo{year}{2020}), ISSN \bibinfo{issn}{0036-8075}.

\bibitem[{\citenamefont{Degen et~al.}(2017)\citenamefont{Degen, Reinhard, and
  Cappellaro}}]{RevModPhys.89.035002}
\bibinfo{author}{\bibfnamefont{C.~L.} \bibnamefont{Degen}},
  \bibinfo{author}{\bibfnamefont{F.}~\bibnamefont{Reinhard}}, \bibnamefont{and}
  \bibinfo{author}{\bibfnamefont{P.}~\bibnamefont{Cappellaro}},
  \bibinfo{journal}{Rev. Mod. Phys.} \textbf{\bibinfo{volume}{89}},
  \bibinfo{pages}{035002} (\bibinfo{year}{2017}).

\bibitem[{\citenamefont{Caves}(1982)}]{caves1982}
\bibinfo{author}{\bibfnamefont{C.~M.} \bibnamefont{Caves}},
  \bibinfo{journal}{Phys. Rev. D} \textbf{\bibinfo{volume}{26}},
  \bibinfo{pages}{1817} (\bibinfo{year}{1982}).

\bibitem[{\citenamefont{Giovannetti et~al.}(2004)\citenamefont{Giovannetti,
  Lloyd, and Maccone}}]{Giovannetti1330}
\bibinfo{author}{\bibfnamefont{V.}~\bibnamefont{Giovannetti}},
  \bibinfo{author}{\bibfnamefont{S.}~\bibnamefont{Lloyd}}, \bibnamefont{and}
  \bibinfo{author}{\bibfnamefont{L.}~\bibnamefont{Maccone}},
  \bibinfo{journal}{Science} \textbf{\bibinfo{volume}{306}},
  \bibinfo{pages}{1330} (\bibinfo{year}{2004}), ISSN \bibinfo{issn}{0036-8075}.

\bibitem[{\citenamefont{Dorogovtsev and
  Mendes}(2013)}]{dorogovtsev2013evolution}
\bibinfo{author}{\bibfnamefont{S.~N.} \bibnamefont{Dorogovtsev}}
  \bibnamefont{and} \bibinfo{author}{\bibfnamefont{J.~F.}
  \bibnamefont{Mendes}}, \emph{\bibinfo{title}{Evolution of networks: From
  biological nets to the Internet and WWW}} (\bibinfo{publisher}{OUP Oxford},
  \bibinfo{year}{2013}).

\bibitem[{\citenamefont{Barab{\'a}si et~al.}(2016)}]{barabasi2016network}
\bibinfo{author}{\bibfnamefont{A.-L.} \bibnamefont{Barab{\'a}si}}
  \bibnamefont{et~al.}, \emph{\bibinfo{title}{Network science}}
  (\bibinfo{publisher}{Cambridge university press}, \bibinfo{year}{2016}).

\bibitem[{\citenamefont{Cohen et~al.}(2000)\citenamefont{Cohen, Erez, ben
  Avraham, and Havlin}}]{PhysRevLett.85.4626}
\bibinfo{author}{\bibfnamefont{R.}~\bibnamefont{Cohen}},
  \bibinfo{author}{\bibfnamefont{K.}~\bibnamefont{Erez}},
  \bibinfo{author}{\bibfnamefont{D.}~\bibnamefont{ben Avraham}},
  \bibnamefont{and} \bibinfo{author}{\bibfnamefont{S.}~\bibnamefont{Havlin}},
  \bibinfo{journal}{Phys. Rev. Lett.} \textbf{\bibinfo{volume}{85}},
  \bibinfo{pages}{4626} (\bibinfo{year}{2000}).

\bibitem[{\citenamefont{Chen et~al.}(2017)\citenamefont{Chen, Wu, Xia, and
  Zhang}}]{chen2017robustness}
\bibinfo{author}{\bibfnamefont{Z.}~\bibnamefont{Chen}},
  \bibinfo{author}{\bibfnamefont{J.}~\bibnamefont{Wu}},
  \bibinfo{author}{\bibfnamefont{Y.}~\bibnamefont{Xia}}, \bibnamefont{and}
  \bibinfo{author}{\bibfnamefont{X.}~\bibnamefont{Zhang}},
  \bibinfo{journal}{IEEE Transactions on Circuits and Systems II: Express
  Briefs} \textbf{\bibinfo{volume}{65}}, \bibinfo{pages}{115}
  (\bibinfo{year}{2017}).

\bibitem[{\citenamefont{Newman}(2010)}]{newman2018networks}
\bibinfo{author}{\bibfnamefont{M.}~\bibnamefont{Newman}},
  \emph{\bibinfo{title}{Networks — An Introduction}}
  (\bibinfo{publisher}{Oxford university press}, \bibinfo{year}{2010}).

\bibitem[{\citenamefont{Mezard and Montanari}(2009)}]{mezard2009information}
\bibinfo{author}{\bibfnamefont{M.}~\bibnamefont{Mezard}} \bibnamefont{and}
  \bibinfo{author}{\bibfnamefont{A.}~\bibnamefont{Montanari}},
  \emph{\bibinfo{title}{Information, physics, and computation}}
  (\bibinfo{publisher}{Oxford University Press}, \bibinfo{year}{2009}).

\bibitem[{\citenamefont{Newman}(2006)}]{internetdata}
\bibinfo{author}{\bibfnamefont{M.~E.~J.} \bibnamefont{Newman}},
  \bibinfo{journal}{The University of Oregon Route Views Project}
  (\bibinfo{year}{2006}).

\bibitem[{\citenamefont{Munro et~al.}(2015)\citenamefont{Munro, Azuma, Tamaki,
  and Nemoto}}]{munro2015inside}
\bibinfo{author}{\bibfnamefont{W.~J.} \bibnamefont{Munro}},
  \bibinfo{author}{\bibfnamefont{K.}~\bibnamefont{Azuma}},
  \bibinfo{author}{\bibfnamefont{K.}~\bibnamefont{Tamaki}}, \bibnamefont{and}
  \bibinfo{author}{\bibfnamefont{K.}~\bibnamefont{Nemoto}},
  \bibinfo{journal}{IEEE Journal of Selected Topics in Quantum Electronics}
  \textbf{\bibinfo{volume}{21}}, \bibinfo{pages}{78} (\bibinfo{year}{2015}).

\bibitem[{\citenamefont{Bennett et~al.}(1996)\citenamefont{Bennett, Brassard,
  Popescu, Schumacher, Smolin, and Wootters}}]{Bennett1996}
\bibinfo{author}{\bibfnamefont{C.~H.} \bibnamefont{Bennett}},
  \bibinfo{author}{\bibfnamefont{G.}~\bibnamefont{Brassard}},
  \bibinfo{author}{\bibfnamefont{S.}~\bibnamefont{Popescu}},
  \bibinfo{author}{\bibfnamefont{B.}~\bibnamefont{Schumacher}},
  \bibinfo{author}{\bibfnamefont{J.~A.} \bibnamefont{Smolin}},
  \bibnamefont{and} \bibinfo{author}{\bibfnamefont{W.~K.}
  \bibnamefont{Wootters}}, \bibinfo{journal}{Phys. Rev. Lett.}
  \textbf{\bibinfo{volume}{76}}, \bibinfo{pages}{722} (\bibinfo{year}{1996}).

\bibitem[{\citenamefont{Bennett et~al.}(1993)\citenamefont{Bennett, Brassard,
  Cr\'epeau, Jozsa, Peres, and Wootters}}]{Bennett1993}
\bibinfo{author}{\bibfnamefont{C.~H.} \bibnamefont{Bennett}},
  \bibinfo{author}{\bibfnamefont{G.}~\bibnamefont{Brassard}},
  \bibinfo{author}{\bibfnamefont{C.}~\bibnamefont{Cr\'epeau}},
  \bibinfo{author}{\bibfnamefont{R.}~\bibnamefont{Jozsa}},
  \bibinfo{author}{\bibfnamefont{A.}~\bibnamefont{Peres}}, \bibnamefont{and}
  \bibinfo{author}{\bibfnamefont{W.~K.} \bibnamefont{Wootters}},
  \bibinfo{journal}{Phys. Rev. Lett.} \textbf{\bibinfo{volume}{70}},
  \bibinfo{pages}{1895} (\bibinfo{year}{1993}).

\bibitem[{\citenamefont{\ifmmode~\dot{Z}\else \.{Z}\fi{}ukowski
  et~al.}(1993)\citenamefont{\ifmmode~\dot{Z}\else \.{Z}\fi{}ukowski,
  Zeilinger, Horne, and Ekert}}]{PhysRevLett.71.4287}
\bibinfo{author}{\bibfnamefont{M.}~\bibnamefont{\ifmmode~\dot{Z}\else
  \.{Z}\fi{}ukowski}},
  \bibinfo{author}{\bibfnamefont{A.}~\bibnamefont{Zeilinger}},
  \bibinfo{author}{\bibfnamefont{M.~A.} \bibnamefont{Horne}}, \bibnamefont{and}
  \bibinfo{author}{\bibfnamefont{A.~K.} \bibnamefont{Ekert}},
  \bibinfo{journal}{Phys. Rev. Lett.} \textbf{\bibinfo{volume}{71}},
  \bibinfo{pages}{4287} (\bibinfo{year}{1993}).

\bibitem[{\citenamefont{Muralidharan et~al.}(2016)\citenamefont{Muralidharan,
  Li, Kim, L{\"u}tkenhaus, Lukin, and Jiang}}]{Muralidharan2016}
\bibinfo{author}{\bibfnamefont{S.}~\bibnamefont{Muralidharan}},
  \bibinfo{author}{\bibfnamefont{L.}~\bibnamefont{Li}},
  \bibinfo{author}{\bibfnamefont{J.}~\bibnamefont{Kim}},
  \bibinfo{author}{\bibfnamefont{N.}~\bibnamefont{L{\"u}tkenhaus}},
  \bibinfo{author}{\bibfnamefont{M.~D.} \bibnamefont{Lukin}}, \bibnamefont{and}
  \bibinfo{author}{\bibfnamefont{L.}~\bibnamefont{Jiang}},
  \bibinfo{journal}{Scientific Reports} \textbf{\bibinfo{volume}{6}},
  \bibinfo{pages}{20463 } (\bibinfo{year}{2016}), \bibinfo{note}{article}.

\bibitem[{\citenamefont{Wehner et~al.}(2018)\citenamefont{Wehner, Elkouss, and
  Hanson}}]{wehner2018quantum}
\bibinfo{author}{\bibfnamefont{S.}~\bibnamefont{Wehner}},
  \bibinfo{author}{\bibfnamefont{D.}~\bibnamefont{Elkouss}}, \bibnamefont{and}
  \bibinfo{author}{\bibfnamefont{R.}~\bibnamefont{Hanson}},
  \bibinfo{journal}{Science} \textbf{\bibinfo{volume}{362}}
  (\bibinfo{year}{2018}).

\bibitem[{\citenamefont{Sahini and M.}(1984)}]{Sahini1984}
\bibinfo{author}{\bibfnamefont{M.}~\bibnamefont{Sahini}} \bibnamefont{and}
  \bibinfo{author}{\bibfnamefont{S.}~\bibnamefont{M.}},
  \emph{\bibinfo{title}{Applications Of Percolation Theory}}
  (\bibinfo{publisher}{CRC Press}, \bibinfo{year}{1984}).

\bibitem[{\citenamefont{Dorogovtsev et~al.}(2008)\citenamefont{Dorogovtsev,
  Goltsev, and Mendes}}]{dorogovtsev2008critical}
\bibinfo{author}{\bibfnamefont{S.~N.} \bibnamefont{Dorogovtsev}},
  \bibinfo{author}{\bibfnamefont{A.~V.} \bibnamefont{Goltsev}},
  \bibnamefont{and} \bibinfo{author}{\bibfnamefont{J.~F.}
  \bibnamefont{Mendes}}, \bibinfo{journal}{Rev. Mod. Phys.}
  \textbf{\bibinfo{volume}{80}}, \bibinfo{pages}{1275} (\bibinfo{year}{2008}).

\bibitem[{\citenamefont{Dijkstra}(1959)}]{dijkstra1959note}
\bibinfo{author}{\bibfnamefont{E.~W.} \bibnamefont{Dijkstra}},
  \bibinfo{journal}{Numerische mathematik} \textbf{\bibinfo{volume}{1}},
  \bibinfo{pages}{269} (\bibinfo{year}{1959}).

\bibitem[{\citenamefont{Kurose and Ross}()}]{kurosecomputer}
\bibinfo{author}{\bibfnamefont{J.~F.} \bibnamefont{Kurose}} \bibnamefont{and}
  \bibinfo{author}{\bibfnamefont{K.~W.} \bibnamefont{Ross}},
  \emph{\bibinfo{title}{Computer networking: A top-down approach}}
  (\bibinfo{publisher}{Addison Wesley}, 2017).

\bibitem[{\citenamefont{Tanenbaum and Wetherall}(2011)}]{TanenbaumWetherall11}
\bibinfo{author}{\bibfnamefont{A.~S.} \bibnamefont{Tanenbaum}}
  \bibnamefont{and}
  \bibinfo{author}{\bibfnamefont{D.}~\bibnamefont{Wetherall}},
  \emph{\bibinfo{title}{Computer Networks}} (\bibinfo{publisher}{Prentice
  Hall}, \bibinfo{address}{Boston}, \bibinfo{year}{2011}),
  \bibinfo{edition}{5th} ed., ISBN \bibinfo{isbn}{978-0-13-212695-3}.

\bibitem[{\citenamefont{Erd{\H{o}}s and R{\'e}nyi}(1959)}]{erdds1959random}
\bibinfo{author}{\bibfnamefont{P.}~\bibnamefont{Erd{\H{o}}s}} \bibnamefont{and}
  \bibinfo{author}{\bibfnamefont{A.}~\bibnamefont{R{\'e}nyi}},
  \bibinfo{journal}{Publ. Math. Debrecen} \textbf{\bibinfo{volume}{6}},
  \bibinfo{pages}{290} (\bibinfo{year}{1959}).

\bibitem[{\citenamefont{Bollob{\'{a}}s and Riordan}(2004)}]{Bollobas2004}
\bibinfo{author}{\bibfnamefont{B.}~\bibnamefont{Bollob{\'{a}}s}}
  \bibnamefont{and} \bibinfo{author}{\bibfnamefont{O.~M.}
  \bibnamefont{Riordan}}, \emph{\bibinfo{title}{Mathematical results on
  scale-free random graphs}} (\bibinfo{year}{2004}).

\bibitem[{\citenamefont{Albert et~al.}(2000)\citenamefont{Albert, Jeong, and
  Barab{\'a}si}}]{Albert2000}
\bibinfo{author}{\bibfnamefont{R.}~\bibnamefont{Albert}},
  \bibinfo{author}{\bibfnamefont{H.}~\bibnamefont{Jeong}}, \bibnamefont{and}
  \bibinfo{author}{\bibfnamefont{A.-L.} \bibnamefont{Barab{\'a}si}},
  \bibinfo{journal}{Nature} \textbf{\bibinfo{volume}{406}}, \bibinfo{pages}{378
  } (\bibinfo{year}{2000}).

\bibitem[{\citenamefont{Callaway et~al.}(2000)\citenamefont{Callaway, Newman,
  Strogatz, and Watts}}]{PhysRevLett.85.5468}
\bibinfo{author}{\bibfnamefont{D.~S.} \bibnamefont{Callaway}},
  \bibinfo{author}{\bibfnamefont{M.~E.~J.} \bibnamefont{Newman}},
  \bibinfo{author}{\bibfnamefont{S.~H.} \bibnamefont{Strogatz}},
  \bibnamefont{and} \bibinfo{author}{\bibfnamefont{D.~J.} \bibnamefont{Watts}},
  \bibinfo{journal}{Phys. Rev. Lett.} \textbf{\bibinfo{volume}{85}},
  \bibinfo{pages}{5468} (\bibinfo{year}{2000}).

\bibitem[{\citenamefont{Barab{\'{a}}si and Albert}(1999)}]{Barabasi1999}
\bibinfo{author}{\bibfnamefont{A.~L.} \bibnamefont{Barab{\'{a}}si}}
  \bibnamefont{and} \bibinfo{author}{\bibfnamefont{R.}~\bibnamefont{Albert}},
  \bibinfo{journal}{Science} \textbf{\bibinfo{volume}{286}},
  \bibinfo{pages}{509} (\bibinfo{year}{1999}), ISSN \bibinfo{issn}{00368075},
  \eprint{9910332}.

\bibitem[{\citenamefont{Dall and Christensen}(2002)}]{PhysRevE.66.016121}
\bibinfo{author}{\bibfnamefont{J.}~\bibnamefont{Dall}} \bibnamefont{and}
  \bibinfo{author}{\bibfnamefont{M.}~\bibnamefont{Christensen}},
  \bibinfo{journal}{Phys. Rev. E} \textbf{\bibinfo{volume}{66}},
  \bibinfo{pages}{016121} (\bibinfo{year}{2002}).

\bibitem[{\citenamefont{Sasaki et~al.}(2011)\citenamefont{Sasaki, Fujiwara,
  Ishizuka, Klaus, Wakui, Takeoka, Miki, Yamashita, Wang, Tanaka
  et~al.}}]{sasaki2011field}
\bibinfo{author}{\bibfnamefont{M.}~\bibnamefont{Sasaki}},
  \bibinfo{author}{\bibfnamefont{M.}~\bibnamefont{Fujiwara}},
  \bibinfo{author}{\bibfnamefont{H.}~\bibnamefont{Ishizuka}},
  \bibinfo{author}{\bibfnamefont{W.}~\bibnamefont{Klaus}},
  \bibinfo{author}{\bibfnamefont{K.}~\bibnamefont{Wakui}},
  \bibinfo{author}{\bibfnamefont{M.}~\bibnamefont{Takeoka}},
  \bibinfo{author}{\bibfnamefont{S.}~\bibnamefont{Miki}},
  \bibinfo{author}{\bibfnamefont{T.}~\bibnamefont{Yamashita}},
  \bibinfo{author}{\bibfnamefont{Z.}~\bibnamefont{Wang}},
  \bibinfo{author}{\bibfnamefont{A.}~\bibnamefont{Tanaka}},
  \bibnamefont{et~al.}, \bibinfo{journal}{Optics express}
  \textbf{\bibinfo{volume}{19}}, \bibinfo{pages}{10387} (\bibinfo{year}{2011}).

\bibitem[{\citenamefont{Jiang et~al.}(2009)\citenamefont{Jiang, Taylor, Nemoto,
  Munro, Van~Meter, and Lukin}}]{Jiang2009}
\bibinfo{author}{\bibfnamefont{L.}~\bibnamefont{Jiang}},
  \bibinfo{author}{\bibfnamefont{J.~M.} \bibnamefont{Taylor}},
  \bibinfo{author}{\bibfnamefont{K.}~\bibnamefont{Nemoto}},
  \bibinfo{author}{\bibfnamefont{W.~J.} \bibnamefont{Munro}},
  \bibinfo{author}{\bibfnamefont{R.}~\bibnamefont{Van~Meter}},
  \bibnamefont{and} \bibinfo{author}{\bibfnamefont{M.~D.} \bibnamefont{Lukin}},
  \bibinfo{journal}{Phys. Rev. A} \textbf{\bibinfo{volume}{79}},
  \bibinfo{pages}{032325} (\bibinfo{year}{2009}).

\bibitem[{\citenamefont{Munro et~al.}(2012)\citenamefont{Munro, Stephens,
  Devitt, Harrison, and Nemoto}}]{munro2012}
\bibinfo{author}{\bibfnamefont{W.~J.} \bibnamefont{Munro}},
  \bibinfo{author}{\bibfnamefont{A.~M.} \bibnamefont{Stephens}},
  \bibinfo{author}{\bibfnamefont{S.~J.} \bibnamefont{Devitt}},
  \bibinfo{author}{\bibfnamefont{K.~A.} \bibnamefont{Harrison}},
  \bibnamefont{and} \bibinfo{author}{\bibfnamefont{K.}~\bibnamefont{Nemoto}},
  \bibinfo{journal}{Nat. Photonics} \textbf{\bibinfo{volume}{6}},
  \bibinfo{pages}{777} (\bibinfo{year}{2012}).

\bibitem[{\citenamefont{Muralidharan et~al.}(2014)\citenamefont{Muralidharan,
  Kim, L\"utkenhaus, Lukin, and Jiang}}]{Muralidharan2014}
\bibinfo{author}{\bibfnamefont{S.}~\bibnamefont{Muralidharan}},
  \bibinfo{author}{\bibfnamefont{J.}~\bibnamefont{Kim}},
  \bibinfo{author}{\bibfnamefont{N.}~\bibnamefont{L\"utkenhaus}},
  \bibinfo{author}{\bibfnamefont{M.~D.} \bibnamefont{Lukin}}, \bibnamefont{and}
  \bibinfo{author}{\bibfnamefont{L.}~\bibnamefont{Jiang}},
  \bibinfo{journal}{Phys. Rev. Lett.} \textbf{\bibinfo{volume}{112}},
  \bibinfo{pages}{250501} (\bibinfo{year}{2014}).

\bibitem[{\citenamefont{Bomze et~al.}(1999)\citenamefont{Bomze, Budinich,
  Pardalos, and Pelillo}}]{bomze1999maximum}
\bibinfo{author}{\bibfnamefont{I.~M.} \bibnamefont{Bomze}},
  \bibinfo{author}{\bibfnamefont{M.}~\bibnamefont{Budinich}},
  \bibinfo{author}{\bibfnamefont{P.~M.} \bibnamefont{Pardalos}},
  \bibnamefont{and} \bibinfo{author}{\bibfnamefont{M.}~\bibnamefont{Pelillo}},
  \emph{\bibinfo{title}{The maximum clique problem}}
  (\bibinfo{publisher}{Springer}, \bibinfo{year}{1999}).

\bibitem[{\citenamefont{Karp}(1972)}]{karp1972reducibility}
\bibinfo{author}{\bibfnamefont{R.~M.} \bibnamefont{Karp}}, in
  \emph{\bibinfo{booktitle}{Complexity of computer computations}}
  (\bibinfo{publisher}{Springer}, \bibinfo{year}{1972}), pp.
  \bibinfo{pages}{85--103}.

\bibitem[{\citenamefont{Riordan and Wormald}(2010)}]{riordan2010diameter}
\bibinfo{author}{\bibfnamefont{O.}~\bibnamefont{Riordan}} \bibnamefont{and}
  \bibinfo{author}{\bibfnamefont{N.}~\bibnamefont{Wormald}},
  \bibinfo{journal}{Combinatorics, Probability and Computing}
  \textbf{\bibinfo{volume}{19}}, \bibinfo{pages}{835} (\bibinfo{year}{2010}).

\end{thebibliography}
\end{document}